\documentclass[aps,twocolumn,showpacs]{revtex4}
\usepackage{graphicx}

\begin{document}

\title{Cost of material or information flow in complex transportation networks}

\author{L. A. Barbosa$^{1,\ast}$ and J. K. L. da Silva$^{2,\dagger}$}

\affiliation{
$^{1}$Centro de Forma\c c\~ao de Professores, Universidade Federal do Rec\^oncavo da Bahia, 45.300-000, Amargosa, BA, Brazil \\
$^{2}$ Departamento de F\'\i sica, Instituto de Ci\^encias Exatas, Universidade Federal de Minas Gerais\\
C. P. 702, 30161-970, Belo Horizonte, MG, Brazil\\}

\date{\today}\widetext


\begin{abstract}

To analyze the transport of information or material from a source to every node of a  network we use two quantities introduced in the study of river networks: the cost and the flow. For a network with $K$ nodes and $M$ levels, we show that an upper bound to the global cost is $C_{0,max}\propto KM$. From numerical simulations for spanning tree networks with scale-free topology  and with $10^2$ up to $10^7$ nodes,
it is found, for large $K$, that the average number of levels and the global cost are given by 
$M\propto \ln(K)$ and  $C_0\propto K\ln (K)$, respectively. These results agree very well with the ones
obtained from a mean-field approach.
If the network is
characterized by a degree distribution of connectivity $P(k)\propto k^{- \gamma}$, we also find
that the transport efficiency increases as long as $\gamma$ decreases and that spanning tree networks with scale-free topology are more 
optimized to transfer information or material than random networks.

\end{abstract}

\maketitle

There is a great interest in the study of networks and its applications 
\cite{mendes-02,barabasi-02,newman-03} because complex network structure underlies many biological, social and technological systems. An interesting development was the discovery that for many real networks, such as Internet \cite{faloutsos-99}, metabolic
networks \cite{jeong-00} and citation networks \cite{redner-98}, the degree distribution follows a long-tailed power law relationship $P(k)\propto k^{- \gamma}$, where $k$ is the number of links of a node \cite{barabasi-99}.  On the other hand, the degree distribution of a random graph is given by a Poisson distribution with a peak at $P(\langle k \rangle)$. Note that in a random graph the edges are placed randomly, implying that the greater part of the nodes have approximately the same degree $\langle k \rangle$ as in the classical Erdos-Renyi model.

The majority of papers about complex networks are focused on static topological properties or on models for their growth. In this paper we are interested in the information, material or energy flow in complex transportation networks. Using tools developed to study transportation networks \cite{rinaldo,banavar-99,garla1-03}, we
study the relation between network topology and efficiency to transfer information, material or energy from a source to every node of the network. Using a simple argument developed for us and a colleague 
\cite{barbosa-pre}, we first determine that an upper bound for the global cost for a  network with $K$ nodes and $M$ levels is 
$C_{0,max}\propto MK$. Then we show for a spanning tree network with scale-free topology that the number of levels grows with the number of nodes as $M\propto \ln(K)$ and that the cost is proportional to $K\ln (K)$. Both results, which are valid in the limit of large $K$, are obtained from numerical simulations for large networks ($K$ varying from $10^2$ up to $10^7$ nodes) and from a mean-field approach. This implies that $C_{0}\propto C_{0,max}\propto MK$ for a  spanning tree network with scale-free topology.

 In general, the analysis of network topology is done by using concepts such as small world character \cite{watts-98}, clustering coefficient \cite{watts-98} and degree distribution \cite{barabasi-99}. However, transport properties have not been extensively studied. Lopez et al. \cite{lopez} have investigated the transport properties of scale-free and Erdos-Renyi networks by analyzing the conductance $G$ between two arbitrarily chosen nodes. They showed that for scale-free networks with $\gamma \geq 2$ the conductance display a power law tail distribution $\phi_{SF}(G) \propto G^{-g_G}$, that is related to the degree distribution $P(k)$ by $g_G=2 \gamma -1$. On the other hand, the conductivity distribution of Erdos-Renyi networks $\phi_{ER}(G)$ decays exponentially.   
The authors concluded that transport in scale-free networks is better than in Erdos-Renyi ones. Castro e Silva et al. \cite{alcides} have studied a deterministic Boolean dynamics on scale-free networks using a damage spreading technique. They showed that the Hamming distance and the number of 1's exhibit power law behavior, with the exponents depending on the value of $\gamma$.

Since the optimization of information (material) flow in networks is an issue of great importance in diverse disciplines \cite{lefeber-06}, it is natural to ask which network topology is more efficient to transfer information (or material) from a source to every node of the network. To answer this question, we use tools developed to study transportation networks \cite{rinaldo,banavar-99,garla1-03}. A transportation system is composed by a source (a central node) and a set of $K$ nodes to be reached. Each of the nodes is connected to one or more of its neighbors nodes in such way that there is a route from the source to every node of the network. This characterizes a spanning
network that can present loops. Since any transportation network has, in general,  a main route from the source to every node, we can
define a spanning tree of the network as a loop less subset of the network in which each node can be reached from the source by the flow
\cite{rinaldo,banavar-99,garla1-03}. Here, let us consider only spanning tree networks, in which the flow rate does not change with time and that the source has an outward flow whose exactly equals the sum of the flow rate into all the links. Examples of such networks include the vascular system of mammals \cite{new-metab}, electric circuits and river drainage basins \cite{rinaldo}.

To generate the spanning tree networks with scale-free topology, we use a growing method with re-direction of links \cite{redner-02}. We start with two nodes, the source $X$ and a node $Y$. The node $X$ or the source is its own ancestor and the ancestor of node $Y$. To insert a new node $Z$, we first select  a node of the network with equal probability. With probability $1-r$ the link is established between the node $Z$ and the selected node, and with probability $r$, the link is established with the ancestor of the selected node.
When we have $r=0$ (no re-direction), a new node $Z$ can be directly connected to each node of the network with equal probability. This constitutes a random network with Poisson degree distribution of connectivity. For values of $r$ in the interval $[0, 1]$, the network has a long-tailed scale-free distribution of connectivity with different values of $\gamma$. Since the relation between $r$ and $\gamma$ is given by $\gamma = 1+\frac{1}{r}$, the case $r=0.5$ corresponds to the linear preferential attachment ($\gamma=3$) \cite{barabasi-99}. 
If $r=1$, the link is always established with the ancestor of the selected node and the network has a ``star-like'' configuration, in which all the nodes are connected directly to the source. In this case, the source has $K$ links and the others $K$ nodes have only one link. The number of levels $M$ of a network can be defined as the number of links among the source and the most distant node. Then, the ``star-like'' network has only one level, $M=1$. Note that the networks built with this method have a tree-like structure, without links between nodes of the same level. Thus, each node is connected to the source by only one path.

Let us measure the cost of transport in these networks using two quantities of the river network theory \cite{rinaldo} and  food-web
studies \cite{garla1-03}. The first one is the quantity $A_i$ of nodes constituting the sub-tree that begins in the node $i$, plus the node $i$ itself. The other quantity is the transportation cost, defined by the sum $C_i=\sum_{k}A_{k}$, where $k$ runs over the sub tree that begins in the node $i$. In analogy with river networks, $A_i$ can be related to the water flux arriving in site $i$ and $C_i$ is identified as the volume of water contained in the sub tree that begins in $i$. Labeling the source by $i=0$, we have that $A_0 = K+1$, $K$ nodes plus the source, is the total size of the network and $C_0$ is the global cost of the transportation system.
For instance, let us consider two networks arrangement, the ``star-like'' and the ``chain-like'' configurations. In the first case the source is at the center and all the nodes are directly connected to it, while in the second case, the ``chain-like'' networks, all nodes have only one incoming link and one outgoing link, except the source and the most distant node. It is knowledge that these configurations are the extreme cases in sense of cost of transport from the source to every node in the network. If there is no constraint on the topology of the network, the spanning tree can be ``chain-like'', ``star-like'' or something in between. Note that the ``star-like'' configuration is the most efficient case and the ``chain-like'' is the less efficient one. Any other tree like structure is between these limits. Motivated by the works of Banavar et al. \cite{banavar-99} and West et al. \cite{west-97}, Garlaschelli \cite{garla1-03} supposed that in general $C_0$ and $K$ are related by a power law relationship $C_0 \propto K^{\eta}$, where $\eta$ quantifies the degree of optimization of the transportation system. It is easy to show that for the ``star-like''  and `chain-like'' configurations  
we have that $\eta=1$ and $\eta=2$, respectively.

In a previous paper, we and a colleague \cite{barbosa-pre}  derived general relations for the upper and the lower bounds for the global cost $C_0$ in a general spanning tree network. Our argument is based in the special characteristic of the cost function: a node put as near as possible of the source has a minimum contribution to the global cost; and a node put as far as possible of the source has a maximum contribution to the global cost. So, we can understand intuitively why the ``star-like'' and ``chain-like'' configurations have the minimum and maximum value of the cost.
Let us consider a spanning tree network with $M$ levels and $K$ nodes. To obey the constraint of $M$ levels, we put one node in each level. Now we must distributed the reminder $K-M$ nodes. If we put all the reminder nodes directly linked in the source we have the network with the minimum value of the cost of transport. In other hand, if we put the reminder nodes in the last level we have the network with the maximum transportation cost. Therefore the minimum and the maximum cost are given by
\begin{eqnarray*}
C_{0_{mim}}&=&1+2K+\frac{M}{2}(M-1)~~,\\
C_{0_{max}}&=&1+K(M+1)+\frac{M}{2}(1-M)~~.
\end{eqnarray*}
In the limit that $K>>M$ and $M>>1$, the above equations reduces to
\begin{equation}
C_{0_{mim}}\propto K~~,~{\rm and}~~~
C_{0_{max}}\propto KM~~.
\label{eqKM}
\end{equation}

Let us look at the two extreme configurations again. The mean number of levels in the ``star-like'' and ``chain-like'' networks scales as $M_{sl} \propto K^0$ and $M_{cl} \propto K^1$, respectively. In the first case, using the above equations, the upper and lower bounds of the exponent $\eta$ have the same
value $\eta_{max}=\eta_{min}=1$. In the ``chain-like'' case, $\eta$
seems to be bounded between $1$ e $2$. Note that $\eta_{min}=1$ is valid  for all networks because
 the minimum cost is fixed. On the other hand, the maximum cost can vary if  $M$ depends on $K$.
 For a network with a topology between the two extreme cases, we expect that $M$ depends on $K$ and
 that the cost be proportional to the maximum cost ($C_0\propto C_{0,max}\propto KM$).

\begin{figure}[hbt]
\centerline{\includegraphics[width=8.5cm,height=7.5cm]{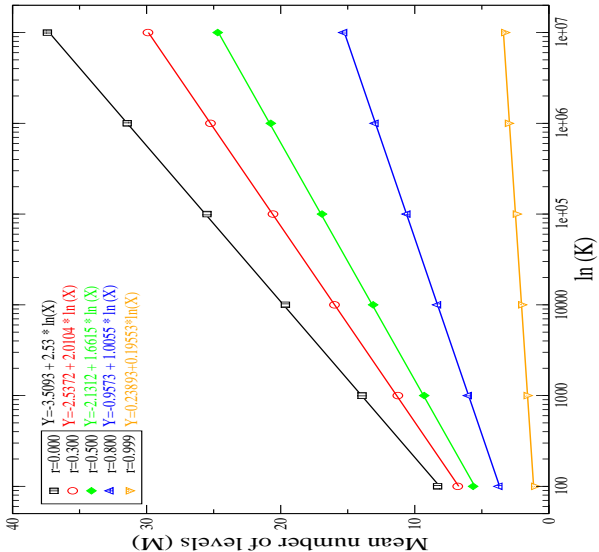}}
\caption{"(Color online)"  Linear-Log plots of the mean number of levels versus the network size $K$
for spanning tree networks with scale-free topology. The networks are built with different  re-direction 
probability $r$.} \label{figMK}
\end{figure}

Now, let us return to  the case of scale-free networks grown by the re-direction of links \cite{redner-02}. In figure (\ref{figMK}) we have plotted the mean number of levels $M$ against the network size $K$ for different values of $r$. Observe that in all cases we have that $M \propto \ln K$. This simulation results suggest that in these networks the cost function is given by

\begin{equation}
C_0 \propto K \ln K~~.\label{eq-jaff}
\end{equation}

This result is in contrast with the premise that the shape of $C_0$ as a function of $K$ always follows a power-law relation $C_0 \propto K^{\eta}$, where the scaling exponent $\eta$ quantifies the degree of optimization of the transportation system used in the study of food webs 
\cite{garla1-03,barbosa-pre,Camacho,garla3}. It is worth mentioning that in this problem, the number
of levels $M$ does not grow with $K$ and, usually is assumed constant ($M < 6$). This implies that
$C_0\propto K$ ($\eta=1$) for food webs because the upper and lower bounds have the same scaling behavior.

\begin{figure}[h]
\centerline{\includegraphics[width=8.5cm,height=7.5cm]{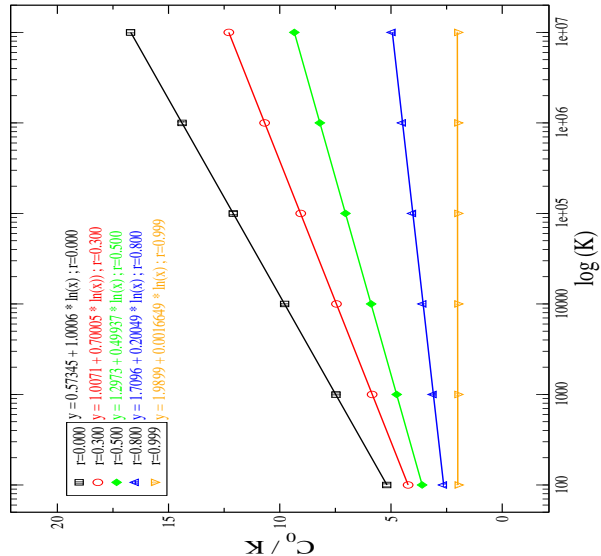}}
\caption{"(Color online)"  Linear-log Plots of the $C_0/K$ versus $\ln K$ for  spanning tree  networks with scale-
free topology. The networks are built with different  re-direction probability $r$. The cost of transport decreases as long as $r$ grows.} \label{fig2}
\end{figure}

In order to verify if Eq. (\ref{eq-jaff}) holds for the scale-free topology, we evaluate numerically the global cost $C_0$ for networks with the size $K$ varying from $10^2$ up to $10^7$. For each network size with a fixed value of the probability of re-direction $r$, we compute the mean
value of the global cost $C_0(K)$ averaged in $10^4$ samples.
In figure \ref{fig2} it is shown the $C_0$ versus $K$ plot for
different values of $r$. Note that cost of transport decreases as long as $r$ grows.
This suggests that among the scale-free
transportation networks the one that has the smallest value of
$\gamma$ is the more efficient. Note that the network constructed
with the re-direction probability $r=0$ (random network) has the biggest cost. 
Observe also that the network with the smallest cost of transport has the smallest
number of levels $M$. Of course, for  networks with the same size $K$, the ones with small number
of levels $M$ are closer to the ``star-like'' configuration than
to the ``chain-like'' one. In fact, the results shown in Fig. \ref{fig2} suggest that in the limit
$r\to 1$ the slope of the line approaches $0$, implying that $C_0\propto K$. In this case the
network has a ''star-like'' shape because we have only a hub with all nodes connected to it.

\vspace{0.2cm}

The numerical results can be explained by a mean-field like approximation.
Let us evaluate the average number of levels, $M$, and the cost, $C_0$, for a scaling-free tree (without loops) with
$K$ nodes. Since $\langle k\rangle$ is the average connectivity, then the source
(node at level 0), is connected, in average, to $\langle k\rangle$ nodes. Each one of the nodes directly connected
to the source (nodes at level 1), are, in average connected to $\langle k\rangle$ other nodes (nodes at level 2).
This argument is repeated until the level $N$, the last level, in such way that the average number of levels $M$  is given by
$$
M={ \langle k\rangle 1 + \langle k\rangle^2 2 + \ldots + \langle k\rangle^N N\over 1+ \langle k\rangle  + \langle k\rangle^2  
\ldots + \langle k\rangle^N}~~.
$$
If we define $S_N=\sum_{n=0}^N \langle k\rangle^n$, the above equation can be written as
$
M=\frac{\langle k\rangle}{S_N}\frac{dS_N}{d\langle k\rangle}.
$
Since the geometric series can be easily evaluated, we obtain that
$$
M= {(N+1)\langle k\rangle^{N+1}\over\langle k\rangle^{N+1} -1} - {\langle k\rangle\over \langle k\rangle -1}~~,
$$
implying that for large $N$ and $\langle k\rangle > 1$ that $M\propto N$. 
On the other hand, $S_N$ must be equal to the number of nodes $K$, namely
$$
S_N={\langle k\rangle^{N+1}-1\over \langle k\rangle}=K~~.
$$
For  $N$ large, we have that $\langle k\rangle^N\propto K$, implying that
\begin{equation}
N\propto  {\ln K\over \ln \langle k\rangle}~~.
\label{eqN}
\end{equation}
Therefore the average number of levels behaves as
$
M\propto \ln K / \ln \langle k\rangle,
$
and agrees very well with the numerical results (see Fig. \ref{figMK}).
We know that $A_i$ is equal to the number of nodes connected to node $i$
plus itself. Then, the recurrence equation is given by
$
A_{N-n}=1+\langle k\rangle A_{N-(n-1)}~~,
$
with the initial condition  $A_N=1$. This equation can be solved and
we obtain the solution, namely
\begin{equation}
A_{N-n}=\sum_{i=0}^n \langle k\rangle^i={\langle k\rangle^{n+1}-1\over\langle k\rangle-1} ~~.
\label{eqA}
\end{equation}

The transportation cost is defined by the sum $C_i=\sum_{k}A_{k}$, where $k$ runs over the sub tree that begins in the node $i$. 
It follows that
$$
C_{N-n}=A_{N-n}+\sum_{i=n-1}^0 \langle k\rangle^{n-i} A_{N-i}~~.
$$
Using Eq. (\ref{eqA}), the above expression can be easily evaluated. We find that
$$
C_{N-n}={\langle k\rangle^{n+1}-1\over\langle k\rangle-1}+{n\langle k\rangle^{n+1}\over\langle k\rangle-1}
+{(1-\langle k\rangle^n)\langle k\rangle\over (\langle k\rangle -1)^2}~~.
$$
Putting $n=N$ we obtain the expression for $C_0$, namely
$$
C_0={\langle k\rangle^{N+1}-1\over\langle k\rangle-1}+{N\langle k\rangle^{N+1}\over\langle k\rangle-1}
+{(1-\langle k\rangle^N)\langle k\rangle\over (\langle k\rangle -1)^2}~~.
$$
When $N$ is large and $\langle k\rangle > 1$, the above equation furnishes that
$
C_0 \propto (N\langle k\rangle^{N})/ (\langle k\rangle-1)~~.
$
Since $\langle k\rangle^{N}\propto K$ and $N\propto M$, it follows that $C_0\propto MK$, meaning
that  $C_0$ is proportional to its maximum value, as supposed before (see Eq. (\ref{eqKM})).
Using again Eq. (\ref{eqN}), $C_0$ can be written as
$$
C_0\propto {K \ln K \over (\langle k\rangle -1)\ln \langle k\rangle}~~.
$$
Moreover,  this equation can be expressed as a function of the parameter $r$ 
by using that  $\langle k\rangle=(\gamma -1)/(\gamma - 2)=1/(1-r)$. The result
$$
C_0\propto -[(1-r)K \ln K]/[ r \ln (1-r)]
$$
agrees very well with the numerical one (see Fig. \ref{fig2}).

In summary, we studied the transportation properties of random 
 spanning tree networks and ones with scale-free topology. We verified that among the scale-free networks, the
efficiency of the transportation properties is related with the
value of the exponent $\gamma$. Compact networks, with small value
of the $\gamma$, are more efficient that non-compact networks,
with large value of $\gamma$. Moreover, we verified also that any 
spanning tree network with scale-free topology is more efficient than random ones,
although  the global cost of both kinds of networks have a similar dependency ($C_0\propto K\ln(K)$)
in the network number of nodes $K$.
\noindent LAB thanks FAPESP and CAPES, Brazilian agencies, for partial financial support. JKL thanks CNPq and FAPEMIG for  financial support. $^{\ast}$ Electronic address: lauro@ufrb.edu.br;
$^{\dagger}$   Electronic address: jaff@fisica.ufmg.br.

\end{document}